# Power Spectral Density Analysis and Correlation of Growth and Morphology of Ni Films on Si Substate

Harsh Bhatt[1,2,*] and Lavanya Negi[3]

[1]Solid State Physics Division, Bhabha Atomic Research Centre Mumbai 400085, India
[2]Homi Bhabha National Institute Anushaktinagar, Mumbai 400094, India
[3]Delhi Public School, Palm Beach Marg Sector – 52 Nerul, Navi Mumbai  400706, India

*harshbhatt@barc.gov.in

**Abstract:**

Ni thin films grown by thermal evaporation and sputtering under different deposition conditions are characterized for structural and morphological properties using X-ray diffraction (XRD) and atomic force microscopy (AFM) techniques. XRD results suggested the growth of polycrystalline *fcc* Ni phase for all the samples. Morphological characteristics of the films were compared by analysing AFM data for root mean square roughness, height-height correlation function and power spectral density (PSD) measurements.  Applying fractal and *k*-correlation fitting models to the PSD data, different morphological parameters are quantified. The study suggested that Ni films grown at higher substrate temperature (~ 150 ºC) by thermal evaporation and at low Ar pressure (~ 0.2 Pa) by sputtering techniques yielded films of small surface roughness with Brownian fractal self-affine surfaces.

**Introduction:**

The study related to the structure, magnetic and catalysis properties of 3d transition metals like iron, nickel and cobalt, have continued beyond the traditional bulk materials and thin films [1-3]. The structure and magnetism of 3d transition metals in the dense phase (under high pressure) are important due to their existence in the Earth's core and thus are still gaining a great deal of attention [4]. While in the bulk form, the dense phases of $3d$ transition metals are obtained by applying external pressure. The growth and nucleation of materials in the form of thin films may show unusual structures that may be obtained in bulk form only under extreme conditions [5]. Recent investigations on Co thin film grown by sputtering technique reported the formation of high-density Co phase at the interfaces, which was found to be nonmagnetic [6,7]. The formation of the high-density Co phase was highly dependent on the structure and morphology of the substrates [7]. The morphology of thin films and their characterization is of prime importance for a better understanding of the physical mechanism of thin film growth and nucleation. Additionally, growing thin films with specifically tailored morphology is critical for acquiring unique physical and chemical properties, e.g. morphology of transition metal film, as a catalyst, is critical for growing carbon nanotubes [8]. Surface morphology also plays an important role in device fabrication for gas sensors, field-effect transistors, and photovoltaic devices [9, 10].

In general, the morphology and microroughness/surface roughness directly affects the functional properties of thin films and thus it is essential to study their surface morphology and microstructure. Experimentally the morphology of a thin film surface can be characterized by measuring the height distribution with a high vertical and spatial resolution using an atomic force microscope (AFM) [11-13]. Surface roughness is usually characterized by the root mean square (RMS) roughness ($\sigma$) that determines the height distribution profile (peak and valleys profile), however, it is not enough to fairly depict the nature of the surface due to a lack of lateral information. Instead, a precise characterization of morphological properties of thin films is carried out by fractal analyses of the roughness using a power spectral density (PSD) function based on the fast Fourier transform (FFT) to determine the dominant spatial frequencies in a surface [11, 12]. The morphological properties of Ni thin films grown by three different methods, viz. thermal evaporation, r. f. sputtering, and electrodeposition, were studied earlier and suggested a drastically different morphology which depends on different nucleation processes involved in growth techniques [13]. However, different growth parameters of the same technique like, Ar pressure in sputtering and substrate temperature in the thermal evaporation technique, may also affect the surface morphology very differently due to different rates of growth and nucleation mechanisms. Here we have studied the structure and morphology of Ni films grown on Si substrate by thermal evaporation and sputtering techniques with different growth parameters. For thermal evaporated and sputtered Ni films, we have varied the substrate temperature and Ar pressure, respectively. While the structure of Ni films grown by both techniques suggested a polycrystalline textured growth. AFM data suggested different surface morphology for different films. Ni films, grown at higher substrate temperature (for thermal annealing) and reduced Ar pressure (for sputtering) yielded smaller surface roughness and a self-affine Brownian fractal surface.

**Experimental:**

*Sample preparation and structural characterization:*

Several thin film samples of Ni with a thickness of ~ 30 nm have been deposited on (100) Si substrate using thermal evaporation and radio frequency (RF) magnetron sputtering techniques with different growth parameters. A typical schematic of the structure of these films is represented in the inset of Fig. 1. The vacuum chambers used for growing Ni films by these two techniques were evacuated to a pressure of $10^{-6}$ bar before the deposition of the films. Si (100) substrates were first cleaned for both organic and inorganic contaminations. Three Ni films were grown by thermal evaporation by keeping all the growth parameters (power, distance between target and substrate etc.) constant except the temperature of the substrate, which was varied for three samples. The substrate temperature for the growth of three Ni films by thermal evaporation, henceforth known as TS1, TS2 and TS3, were kept at RT (~ 25 °C), 100 °C and 150 °C, respectively. In the case of DC magnetron sputtering growth, only the argon (Ar) pressure was varied and we have used three different Ar pressures of 0.8 Pa, 0.4 Pa and 0.2 Pa for growing three Ni films, henceforth known as SS1, SS2 and SS3, respectively. The different growth conditions and related samples are given in Table 1. The films were characterized for structural properties by X-ray diffraction (XRD) using Cu $K_\alpha$ X-ray (wavelength 1.54 Å).

*AFM measurements and power spectral density by fast Fourier transform*

The surface topographical measurements of Ni thin films were investigated using an NT-NDT's Solver P-47 H ambient-based multimode AFM instrument. The images were taken in a non-contact mode with a $Si_3N_4$ tip at room temperature (RT). The cantilever has a spring constant of 0.1 N/m. AFM topographic images were recorded for each film over a scan area of 2 μm × 2 μm with a resolution of 256 × 256 data points. The PSD analysis provides a more complete roughness description by AFM measurements because it intrinsically accounts for the statistical correlation between the different surface heights. The one-dimensional PSD profile by the FFT algorithm was calculated from the scan line profile measurement of AFM image data for Ni films [9, 10, 12]. The computation of the one-dimension discrete PSD profile adopted in this paper is given by [9, 10, 12]:

$$PSD\ (f) = \frac{1}{L}[\sum_{m=1}^{N} Z_m \exp(-2\pi i\ \Delta L\ f\ m)\ \Delta L]^2 \qquad (1)$$

where $L^2$ is the scanned surface area, $N$ is the number of data points per line (x-direction)/row ($N = 256$), $Z_m$ is the height at position m along the x-axis for fixed row number, $f$ is the spatial frequency in the x directions and $\Delta L$ is the sampling distance ($\Delta L = L/N$). The one-dimensional PSD for an AFM image discussed in this paper is averaged over an entire number of rows ($N = 256$). The spatial frequency ($f$) can take a discrete range of values from $\frac{1}{L}, \frac{2}{L}, \frac{3}{L}, \ldots\ldots, \frac{N}{2L}$. The maximum frequency ($\frac{N}{2L}$) is limited by the sampling theorem of Nyquist frequency [10, 13]. We have recorded the AFM image with a scan area of 2.0 μm × 2.0 μm, which gives a sampling rate/distance ($L/N = 3/256$) of 7.8 nm. The minimum and maximum spatial frequency is limited to $1.0/(2.0\ \mu m) \sim 0.5\ \mu m^{-1}$ and $1.0/(2 \times 7.8\ nm) \sim 64.1\ \mu m^{-1}$, which makes the upper and lower-bandwidth limitation of the PSD profiles.

**Results and Discussion:**

Figure 1 shows the XRD pattern obtained from Ni thin films TS1 (open circles) and SS1 (open triangles). We carried out XRD measurements from all the samples which showed similar features. The different Bragg peaks from Ni films are indexed in Fig. 1, suggesting the formation of a polycrystalline Ni phase with a face-centred cubic (*fcc*) structure and a lattice parameter of 3.50 Å and 3.48 Å for samples TS1 and SS1, respectively. The lattice parameter for Ni film is smaller than the Ni bulk phase (~ 3.52 Å), suggesting that films are grown with small (0.6% to 1.0% ) out-of-plane compressive strain. It is also evident from the XRD profiles that the intensity of (111) reflection of Ni is stronger as compared to other peaks for both samples, indicating the presence of (111) texture in the polycrystalline film.

Figure 2 shows the AFM images with a scan size of 2 μm × 2 μm and related height distribution profiles of the Ni films (TS1, TS2 and TS3) grown by thermal evaporation. The AFM image for the sample TS1 (grown at RT) is shown in Fig. 2 (a). The height distribution function of this image, as well as height fluctuation along the diagonal of the film (blue dash line), is shown on the right of the AFM image in Fig. 2 (a) [bottom and top panels]. The Gaussian function is fitted ( solid black line) to the height distribution function obtained from the AFM image that shows a small deviation of the Gaussian function indicating a surface having features distributed non-symmetrically around a mean surface profile (i.e. more than one length scale present on the surface). Similarly, Fig. 2(b) and (c) show the AFM images (on the left) and the corresponding height distribution function and height fluctuation (on the right) of the samples TS2 and TS3, respectively. The Gaussian fit to height distribution function of samples TS2 and TS3 (Fig. 2 (b) and (c)) does not show any deviation suggesting a surface having features distributed symmetrically around a mean surface profile and sample TS3 grown at higher substrate temperature shows more symmetric distribution of height-height fluctuation features (reduced FWHM of Gaussian fit). Figure 3 shows the AFM images and the corresponding height-height distribution functions for samples SS1 (Fig. 3(a)), SS2 (Fig. 3(b)) and SS3 (Fig. 3(c)), grown by sputtering. The Gaussian fit to height distribution function of all these sputtered samples suggests the surface has features distributed symmetrically around a mean surface profile. The surface morphology is usually analyzed by measuring the statistical surface roughness, like RMS roughness, average roughness etc and height-height correlation function.

*Statistical Roughness Analysis:*

The height (or depth) fluctuation and its range in lateral dimension are the two main attributes of surface roughness [14]. RMS and average roughness calculation from AFM data are the simplest and most used parameters for surface topography. The RMS roughness ($\sigma$) of the surface of a given area of an AFM image ($N \times N$) is an attractive and simple statistical measure which can summarize the surface roughness by a single value. The $\sigma$ is defined as the square root of the mean value of the squares of the height, Z (*x*, *y*), from the mean value of heights over the whole area of the surface and is given by:

$$\sigma = \sqrt{\frac{1}{N^2}\sum_{i=1}^{N}\sum_{j=1}^{N}[Z(i,j)-\bar{Z}]^2} \qquad (2)$$

Where N, $Z(i,j)$ and $\bar{Z}$ are the number of data points ( ~ 256 in the present case), the value of height at the $(i,j)$ position, and the mean value of the height, respectively. The mean value of the height is defined as: $\bar{Z} = \frac{1}{N^2}\sum_{i=1}^{N}\sum_{j=1}^{N} Z(i,j)$. Using Eqn. (2) we have calculated the RMS roughness of each sample and the RMS roughness value is given in Table 2. It is evident from the RMS roughness results that on increasing the substrate temperature from RT to 150 ºC for samples grown by thermal evaporation (TS1-TS3), the surface becomes smother i.e. the roughness of the surface decreases by a factor of 4 (changes from 24 to 6 Å). The higher value of substrate temperature may be providing additional energy for the redistribution of addend atoms on the substrate and hence increasing the smoothness of the film. For the films grown by sputtering (SS1 – SS3), the roughness of the films grown at higher Ar pressures (SS1 and SS2) show much higher ( ~ 18 Å) as compared to that of Ni film grown with the lowest Ar pressure (13 Å). While the RMS roughness obtained from an AFM image is the simplest way of defining surface topography but its value is highly dependent on the scan size [13,17] and thus cannot fully characterize the surface.

*Height-Height Correlation Function:*

As mentioned above the surface morphology is characterized by height distribution function and it can be simply reported by a single parameter, surface roughness (RMS height). However, this knowledge about surface roughness hardly provides useful information for many applications, especially where it is crucial to know the extent of roughness decomposition into long or short-wavelength modes. The height-difference correlation function, *g* (*r*) defined in Eqn. (3) contains stochastic information on the surface profile in real space, which is another way of analyzing the surface morphology and yields the fractal scaling behaviour of the surface morphology. The height difference correlation function from AFM data can be calculated as [13, 17]:

$$g(r) = \langle [Z(x_2, y_2) - Z(x_1, y_1)]^2 \rangle \text{ with } r = \sqrt{(x_2 - x_1)^2 + (y_2 - y_1)^2} \quad (3)$$

where symbol $\langle \cdots \rangle$ denotes the spatial average over the surface and Z denotes the height of the interface at the lateral position, *r*. In general, thin films grown under nonequilibrium conditions are expected to develop self-affine surfaces [18] and the average height difference between any two points separated by a distance *r* for self-affine surfaces is described by the phenomenological function defined as [18-20]:

$$g(r) = 2\sigma^2[1 - e^{-(\frac{r}{\xi})^{2h}}] \qquad (4)$$

where $\sigma$, $h$, and $\xi$ are the surface roughness, the Hurst parameter, and the correlation length.[18, 19] The Hurst parameter measures the short-range roughness. On fitting the function defined in Eqn. (4) to the height-height correlation data obtained from AFM measurements (using Eqn. (3)) one obtained these morphological parameters for self-affine surfaces. However, $g(r)$ functions quickly reach a saturation value and the correlation is best defined for $r \gg \xi$. For $r \ll \xi$, the $g(r)$ function, defined in Eqn. (4) is believed not a good fit for the correlation data and can overestimate the Hurst parameter [19]. For self-affine surfaces/interfaces, the Hurst parameter describes the texture of roughness of surfaces/interfaces and it can take a value $0 < h < 1$, defining the fractal dimension of the surface to $D = 3 - h$.

Figures 4 (a) and (b) show the $g(r)$ data (symbol) from AFM measurements using Eqn. (3) and corresponding fits (solid lines) using Eqn. (4) for samples grown by thermal evaporation and sputtering, respectively, on a log-log scale. We have used a Python-based code for the calculation and analysis of the height-height correlation functions defined in Eqns. (3) and (4). The three morphological parameters ($\sigma$, $\xi$ and $h$) obtained from the best fit of $g(r)$ data for different samples are given in Table 1 and also compared in Fig. 4 (c), (d) and (e). Both the roughness and the height-height correlation length decrease with an increase in the substrate temperature for the film grown by thermal evaporation. While the Hurst parameter for film grown by thermal evaporation at a substrate temperature of ~ 150 °C was found to be increased by 23% and suggesting a smooth surface. Similarly, on decreasing the pressure of Ar gas for deposition of Ni film by sputtering the morphological parameters, i.e. the roughness, correlation length and the Hurst parameter, decreases.

*PSD Analysis:*

The PSD of the surface morphology is a fundamental tool for describing the statistical properties of surfaces which contains a more complete description than the RMS roughness and provides useful quantitative information of a thin film surface morphology. Figures 5 (a) and (b) show the PSD data calculated using Eqn (1) from AFM images (size: 2 μm × 2 μm shown in Fig. 2 and 3) of samples grown by thermal evaporation (Fig. 2) and sputtering (Fig. 3), respectively, revealing the decrease in spectrum density with an increase in spatial frequency. The PSD profiles are divided into 3 spatial frequency regions (I, II and III), which provide surface characteristics in different frequency ranges. Usually, many peaks are seen at low frequencies (Region I) suggesting a periodic surface. We did not observe the peaks in this range suggesting a randomly rough surface for all the samples in this length scale of ~ 2 μm. The surface of all Ni films grown by different techniques and growth parameters follow inverse power law at a high-frequency range (region III) and have fractal features. The surface morphological parameters can be obtained by fitting the PSD data using various models and One of the standard models is a power-law model (fractal model), which can be applied at a high-frequency region (region III) and defined as [12]:

$$PSD_{fractal}(f) = \frac{K}{f^v} \quad (5)$$

where $K$, $v$, and $f$ are intrinsic surface parameters and known as spectral strength, spectral indices and spatial frequency, respectively. For self-affine surfaces, the fractal dimension $D$ of the surface can be determined using [12]: $D = \frac{7-v}{2}$, if $1 \leq v \leq 3$. The fractal behaviour of the surface, obtained by fitting PSD data using Eqn. (5), does not show complete characterization of surface roughness and another model, called *k*-correlation *or* ABC model was proposed to fit the PSD data for quantification of surface morphology. The ABC model for PSD is given as [11, 12]:

$$PSD_{ABC}(f) = \frac{A}{(1 + B^2 f^2)^{\frac{C+1}{2}}} \quad (6)$$

where $A$, $B$, and $C$ are model parameters and characterize the random rough surface morphology over a larger frequency range including region II (Fig. 5). The ABC model of PSD function (Eqn. (6)) yields a ''knee,'' determined by $B$, which is equal to the correlation length [11, 12]. The parameter $A$ is defined at a low spatial frequency and gives a measure of roughness. At high-frequency values, beyond the knee, the surface is fractal and the PSD function is determined by $C$. We have used a Python-based code for generating the PSD profile from AFM data (using Eqn. 1) as well as for analysis of PSD measurements using different models given in Eqns. (5) and (6).

Using the PSD models (fractal and ABC) defined in Eqns. (5) and (6), we have fitted the PSD data from AFM measurements for all the samples with the resultant PSD fit model ( $PSD_{fit} = PSD_{fractal} + PSD_{ABC}$ ) and the fitted profiles with data are shown in Fig. 6 on a log-log scale. Symbols represent the PSD data from the AFM image and solid lines are the resultant fit to PSd data. We have also plotted the contribution to PSD fit from the fractal model (black dashed lines) and ABC model (red dash-dot lines). The summary of parameters obtained from best fit to PSD data is given in Table 2. While the surfaces of all the samples show a fractal dimension of ~ 2.5, we found a decrease in the value of spectral strength ($K$) on increasing (decreasing) the substrate temperature (Ar pressure) for thermal evaporated (sputtering) samples suggesting strong dependence of $K$ on these growth parameters. The fractal dimension near 2.5 is mostly defined for Brownian fractal [11, 12] surface suggesting all our films show Brownian fractal surface. For the *k*-correlation model, we obtained a consistent decrease in parameter $A$ on increasing (decreasing) the substrate temperature (Ar pressure) for thermal evaporated (sputtered) films. However, parameter $A$ influences the low-frequency PSD data, which is having more noise due to scan size limitation so it is difficult to conclude the dependence of this parameter on surface roughness. The value of parameter $C$ ~ 4.0, which is a measure of the inverse slope of the PSD curve, suggest that films are grown with a nucleation process of surface diffusion [21]. Thus the higher substrate temperature in the thermal evaporation and low Ar pressure in the sputtering technique favours the growth nucleation and relaxation mechanism of surface diffusion and yields good quality thin film growth.

**Conclusion:**

Ni films grown on Si substrate by thermal evaporation and sputtering under different growth conditions were characterized for structural and morphological properties using XRD and AFM measurements. The surface morphology was studied by the PSD of fast Fourier transform algorism using AFM digital image data. XRD measurements showed the polycrystalline growth of the films grown by both techniques. PSD data were fitted considering both the fractal model (high frequency) and the *k*-correlation model. We found small surface roughness for Ni Films grown at higher (~ 150 °C) substrate temperature by thermal evaporation and the PSD analysis suggested that these surfaces showed self-affine fractal surfaces with a nucleation process of surface diffusion. Similar results were obtained for Ni film grown by sputtering technique with small Ar pressure (0.2 Pa). The different deposition parameters (substrate temperatures and Ar pressures) used in these two techniques also influenced the spectral strength ($K$) of the fractal components of the surfaces. Morphological parameters also suggested that the higher substrate temperature for thermal evaporation and low Ar pressure for sputtering techniques favour the Brownian fractal with surface diffusion for nucleation of films with improved surface topological parameters.

**References:**

1.  P. Gambardella, A. Dallmeyer, K. Maiti, M. C. Malagoli, W. Eberhardt, K. Kern & C. Carbone. Nature 416, 301 (2002).
2.  P. Gandeepan, T. Müller, D. Zell, G. Cera, S. Warratz, and L. Ackermann, Chem. Rev. 119, 2192 (2019).
3.  L. S. Lobo and S. A. C. Carabineiro, Nanomaterials 11, 143 (2021).
4.  C. S. Yoo, H. Cynn, P. Söderlind, and V. Iota, Phys. Rev. Lett. 84, 4132 (2000).
5.  C. M. Schneider, P. Bressler, P. Schuster, J. Kirschner, J. J. de Miguel, and R. Miranda, Phys. Rev. Lett. 64, 1059 (1990).
6.  N. Banu, S. Singh, B. Satpati, A. Roy, S. Basu, P. Chakraborty, H. C. P. Movva, V. Lauter & B. N. Dev, Sci. Rep. 7, 41856 (2017).
7.  N. Banu, S. Singh, S. Basu, A. Roy, H. C P Movva, V. Lauter, B. Satpati and B. N. Dev, Nanotechnology 29, 195703 (2018).
8.  X. Jin, J. Lim, Y. Ha, N. H. Kwon, H. Shin, I. Y. Kim, N.-S. Lee, M. H. Kim, H. Kim and S.-J. Hwang, Nanoscale 9, 12416 (2017).
9.  R. D. Yang, T. Gredig, C. N. Colesniuc, J. Park, I. K. Schuller, W. C. Trogler, and A. C. Kummel, Appl. Phys. Lett. 90, 263506 (2007).
10. A. Gusain, S. Singh, A.K. Chauhan, V. Saxena, P. Jha, P. Veerender, A. Singh, P.V. Varde, S. Basu, D.K. Aswal and S.K. Gupta, Chem. Phys. Lett., 646, 6 (2016).
11. M. Senthilkumar, N.K. Sahoo, S. Thakur, R.B. Tokas, Applied Surface Science 252, 1608 (2005).
12. L. Eftekhari, D. Raoufi, M. J. Eshraghi and M. Ghasemi, Semicond. Sci. Technol. 37, 105011 (2022).
13. S. Singh and S. Basu, Surface & Coatings Technology 201, 952 (2006).
14. R. Gavrila, A. Dinescu, D.Mardare, Romanian Journal of Information Science and Technology, 10, 291 (2007).
15. D. Raoufi, Physica B 405, 451 (2010).
16. J. M. Bennett, Meas. Sci. Technol. 3, 1119 (1992).
17. G. Palasantzas, J. Krim, Phys. Rev. Lett. 73, 3564 (1994).
18. F. Family, Physica A 168, 561 (1990).

19. S. K. Sinha, E. B. Sirota, S. Garoff, and H. B. Stanley, Phy. Rev. B 38, 2297 (1988).
20. S. Singh and S. Basu, Surface Science, 600, 493 (2006).
21. D.G. Stearns, P.B. Mirkarim, E. Spiller, Thin Solid Films 446, 37–49 (2004).

Table 1: Nomaneculture of different Ni thin film samples grown by thermal evaporation and direct current (DC) sputtering with varying parameters.

| Thermal evaporation: Ni (30 nm)/Si-substrate | | RF Sputtering Ni (30 nm)/Si-substrate | |
|---|---|---|---|
| Parameter varied (substrate temperature) | Sample | Parameter varied (pressure of Ar gas) | Sample |
| RT (25 º C) | TS1 | 0.8 Pa | SS1 |
| 100 º C | TS2 | 0.4 Pa | SS2 |
| 150 º C | TS3 | 0.2 Pa | SS3 |

Table 2: Different parameters describing a PSD model are fitted to experimental PSD data for Ni thin film samples grown by thermal evaporation and direct current (DC) sputtering

| Samples | | RMS $\sigma$ (Å) | $\xi$ (Å) | $h$ | Fractal | | k-Correlation | | |
|---|---|---|---|---|---|---|---|---|---|
| | | | | | $K$ ($10^{-4}$ nm) | $v$ | $A$ (nm) | $B$ (nm) | $C$ |
| Thermal Evaporation | TS1 | 24 | 1100 | 0.65 | 1.8 | 2.05 | 50 | 120 | 4.5 |
| | TS2 | 16 | 550 | 0.65 | 0.8 | 2.06 | 10 | 100 | 5 |
| | TS3 | 6 | 800 | 0.78 | 0.16 | 2.03 | 2 | 100 | 4 |
| RF Sputtering | SS1 | 17 | 1100 | 0.96 | 0.4 | 2.03 | 54 | 160 | 5 |
| | SS2 | 18 | 650 | 0.96 | 0.4 | 2.06 | 20 | 80 | 5 |
| | SS3 | 13 | 580 | 0.88 | 0.3 | 2.06 | 13 | 80 | 4 |

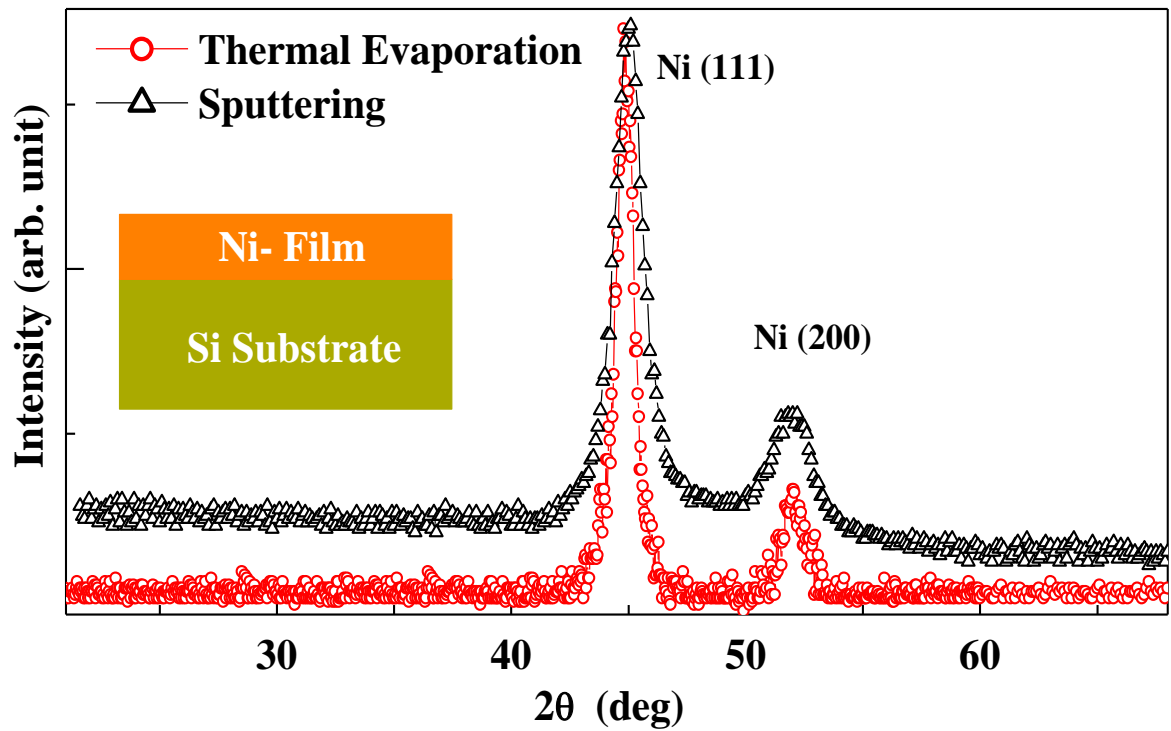

Figure 1: XRD profiles of the Ni film grown by thermal evaporation and sputtering techniques. The (111) and (200) reflections of Ni are indexed in the XRD profiles. Inset show the schematic of the layer structure of Ni film on a Si substrate.

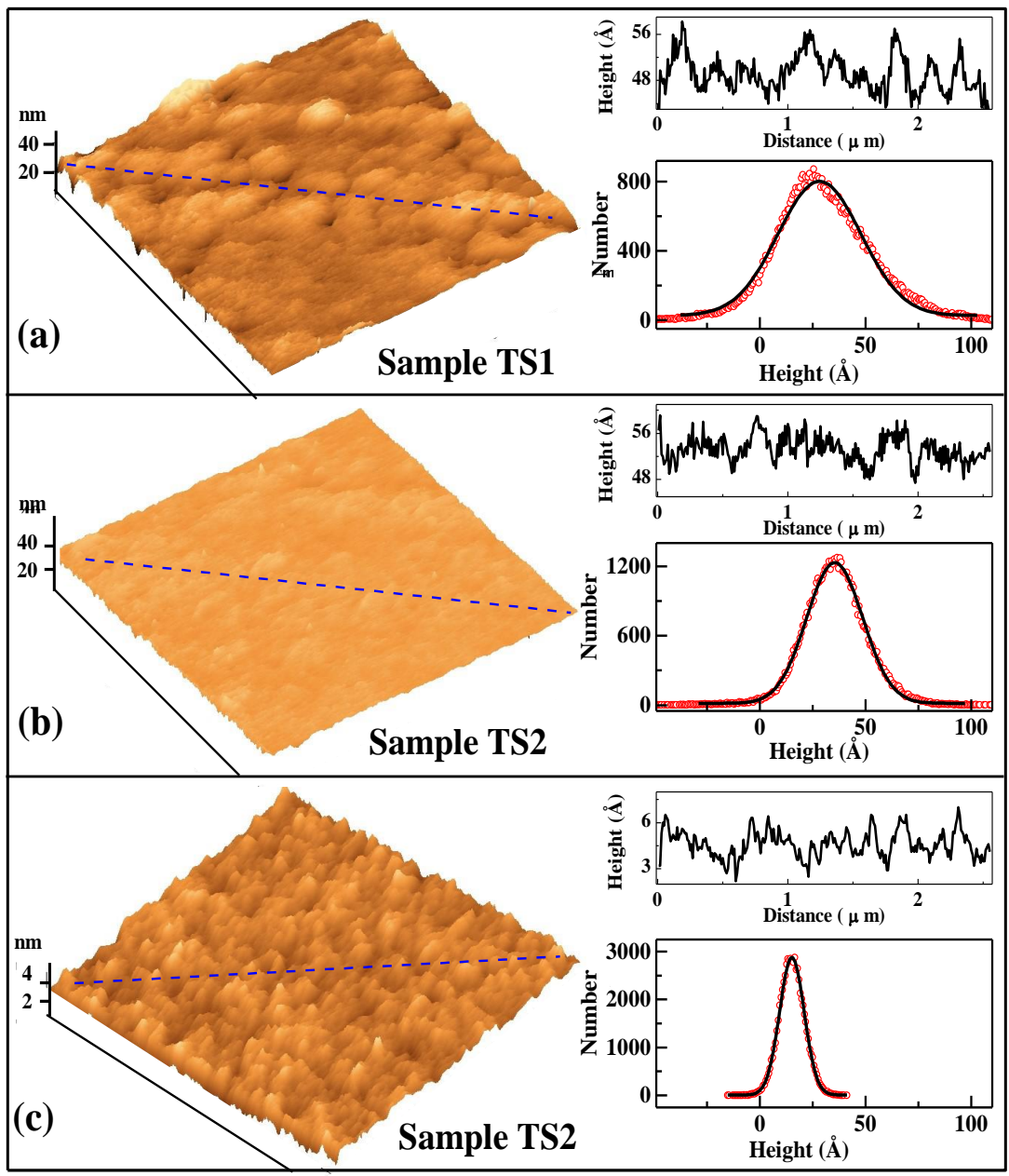

Figure 2: AFM measurements from Ni film grown by thermal evaporation. Surface topography of Ni films, TS1 (a), TS2 (b) and TS3 (c) with a scan area of 2 μm × 2 μm, each. The right panel of each image show the corresponding height distribution (upper) and topological histogram (lower) of the AFM images. The solid curves in the topological histogram are the corresponding Gaussian fits.

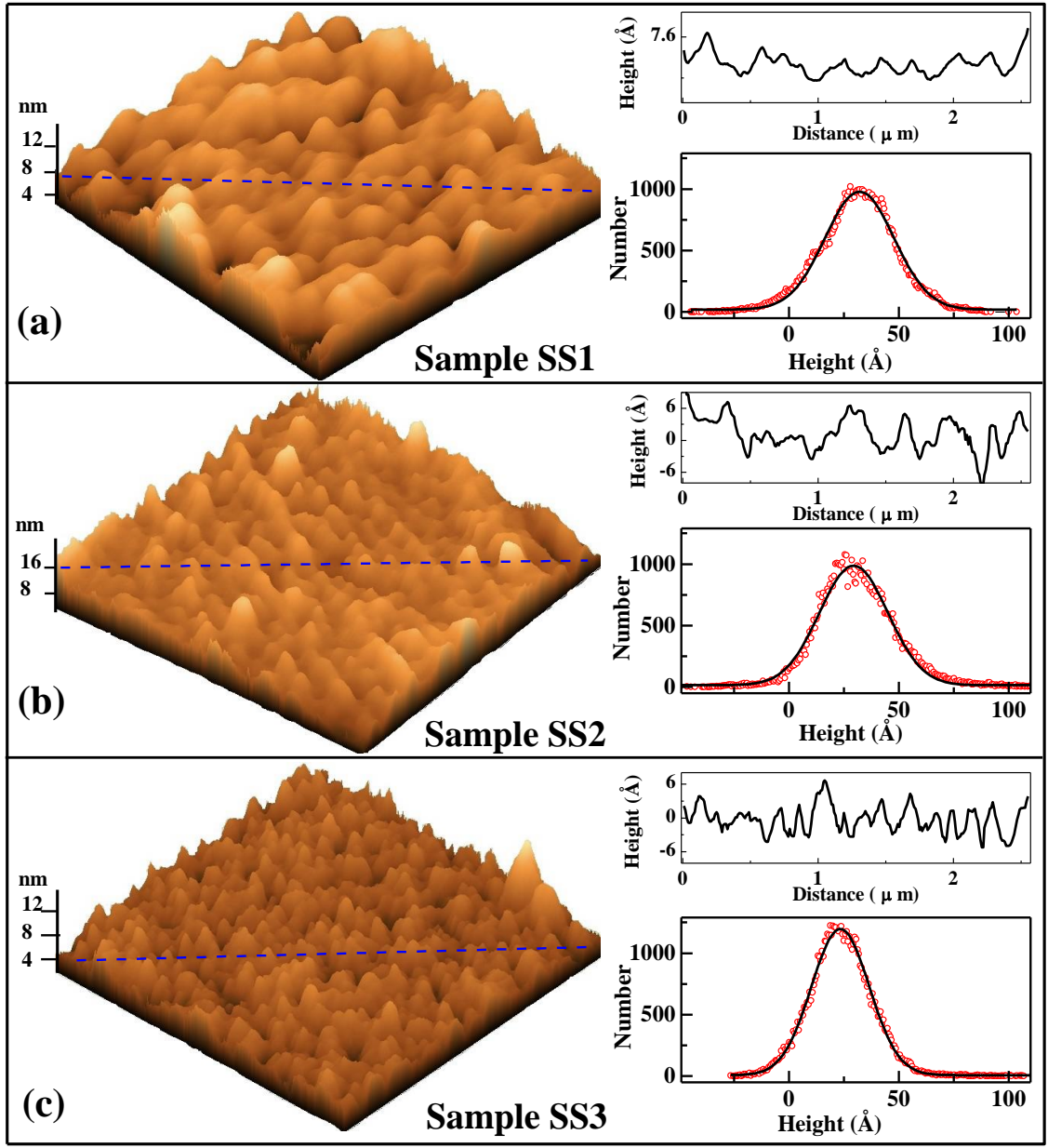

Figure 3: AFM measurements from Ni film grown by sputtering. Surface topography of Ni films, SS1 (a), SS2 (b) and SS3 (c) with a scan area of 2 µm × 2 µm, each. The right panel of each image shows the corresponding height distribution along the diagonal line in the image (upper) and topological histogram (lower) data (red open circles) of the AFM images. The solid curves in the topological histogram are the corresponding Gaussian fits.

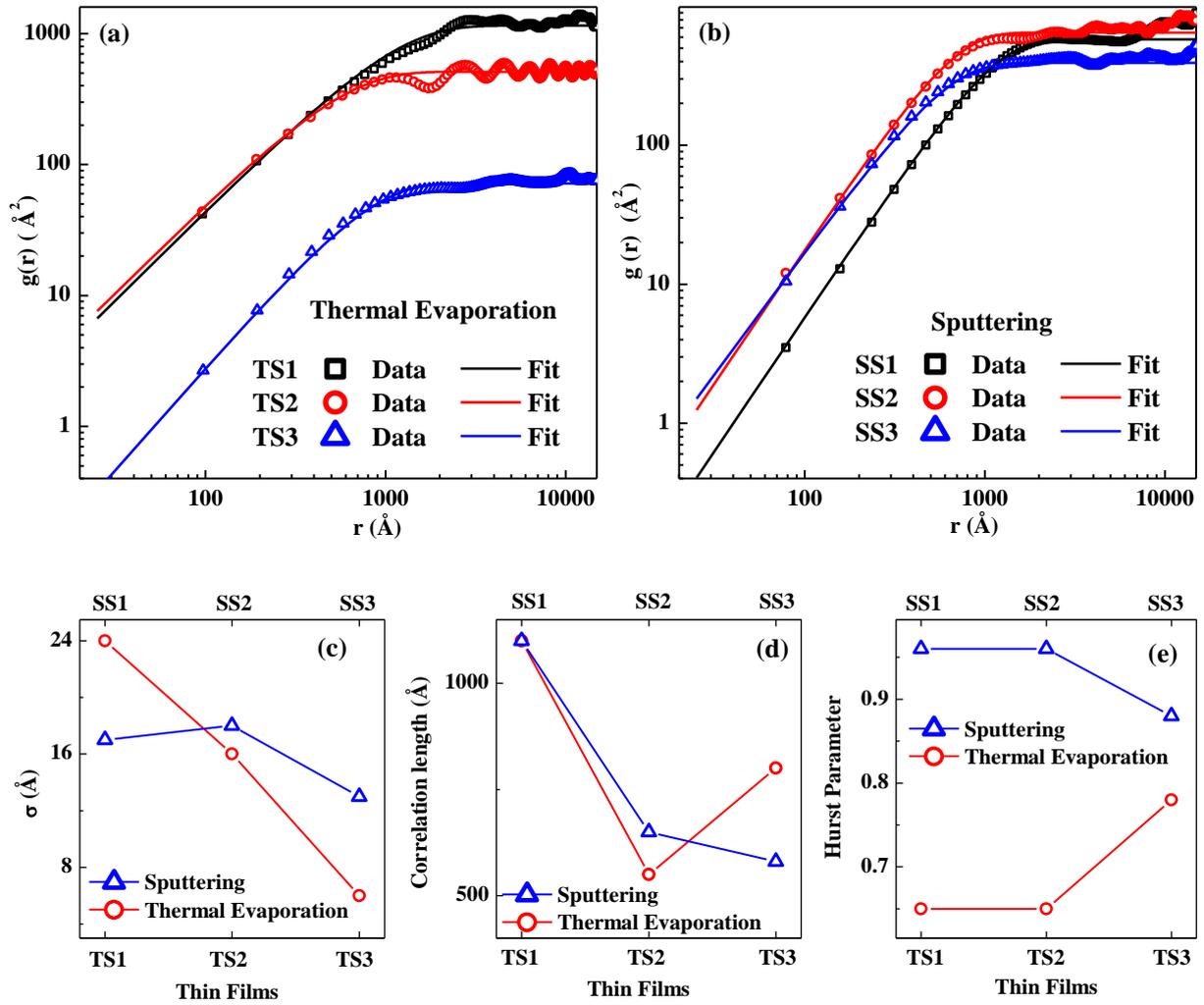

Figure 4: The height difference correlation function, *g* (r), obtained from the AFM data and the corresponding fit for samples grown by (a) thermal evaporation and (b) sputtering. The comparison of morphological parameters, (c) roughness ($\sigma$), (d) correlation lengths ($\xi$) and (e) the Hurst parameter (*h*) for samples grown by two techniques, thermal evaporation (TS1, TS2 and TS3) and sputtering (SS1, SS2, and SS3).

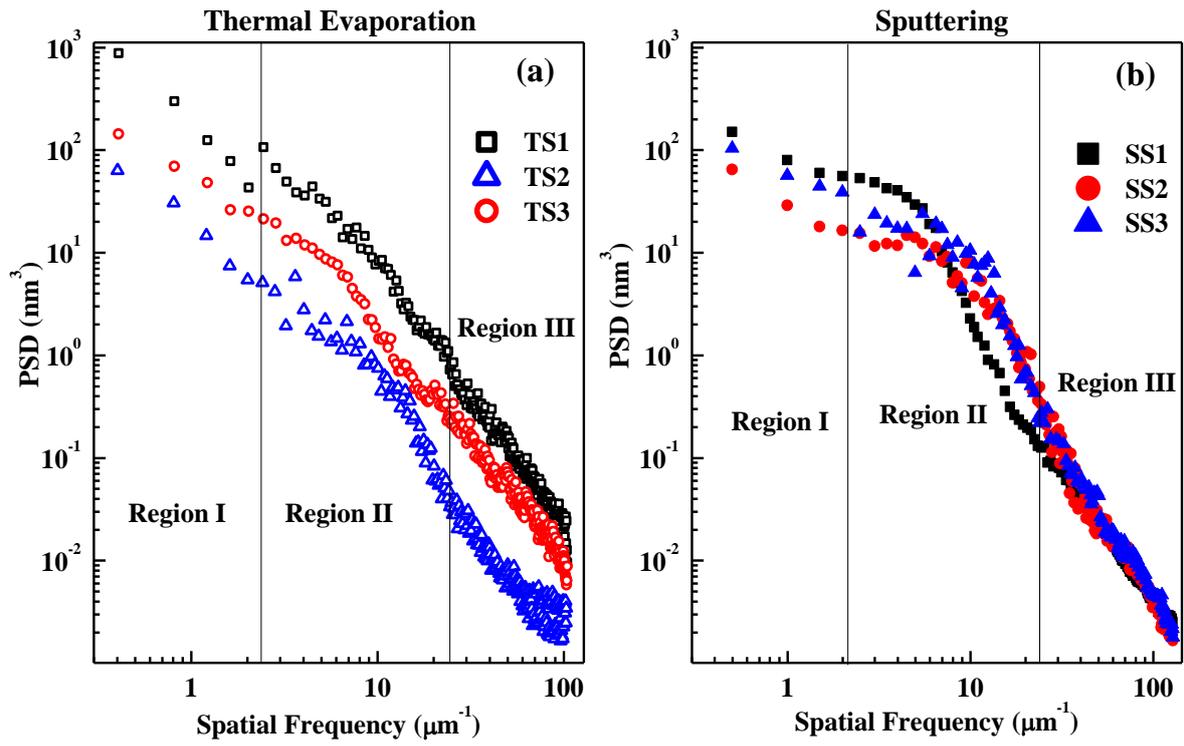

Figure 5: PSD data as a function of spatial frequency for Ni films grown by (a) thermal evaporation and (b) sputtering under different growth conditions. The frequency range is divided into three important regions.

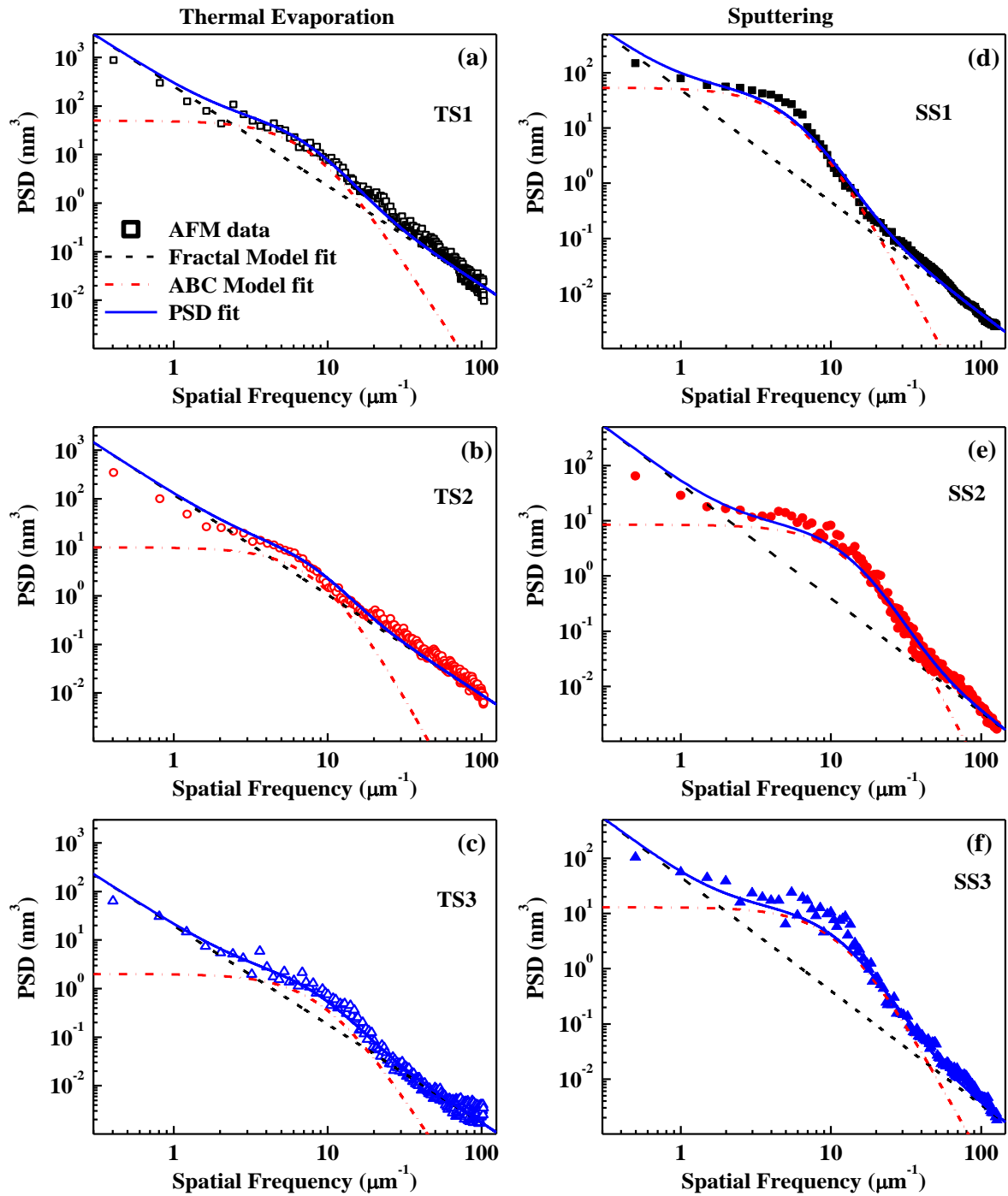

Figure 6: PSD data (symbols) and the corresponding fit considering the contribution from two models (i) fractal model (black dash line) and (ii) k-correlation (ABC) model (red, dash-dot line). The left panel (a to c) and right panel (d to e) show the PSD measurements from Ni film grown by thermal evaporation and sputtering techniques, respectively, under different conditions (see text).